\documentclass[article,pre,10pt,twocolumn,longbibliography]
{revtex4-1}
\usepackage{microtype}

\usepackage[colorlinks,linkcolor=blue,urlcolor=blue,citecolor=blue]{hyperref}

\usepackage{caption}
\usepackage{subcaption}
\usepackage{mathptmx}
\usepackage{times}
\usepackage{epsfig,amsopn}
\usepackage{graphicx}
\usepackage{bm,amsmath,amssymb}
\usepackage{natbib}
\usepackage{braket}
\usepackage{float}
\usepackage{enumerate}
\usepackage{hyperref}
\usepackage{comment}
\allowdisplaybreaks[1]

\DeclareMathOperator{\sech}{sech}
\def\be{\begin{equation}}
\def\ee{\end{equation}}
\def\bse{\begin{subequations}}
\def\ese{\end{subequations}}
\def\bcs{\begin{cases}}
\def\ecs{\end{cases}}
\def\bea{\begin{eqnarray}}
\def\eea{\end{eqnarray}}

\newcommand{\Aref}[1]{Appendix}%

\newcommand{\opunit}{\textrm{1}\kern-0.22em\textrm{l}}

\usepackage{graphicx}

\begin{document}

\title{Three state random energy model}

\author{{\normalsize{}Sumedha$^{1, 2}$}
{\normalsize{}}}
\email{sumedha@niser.ac.in}

\author{{\normalsize{}Matteo Marsili$^{3}$}
{\normalsize{}}}
\email{marsili@ictp.it}

\affiliation{\noindent $^{1}$School of Physical Sciences, National Institute of Science Education and Research, Bhubaneswar, P.O. Jatni, 752050, India}

\affiliation{\noindent $^{2}$Homi Bhabha National Institute, Training School Complex, Anushakti Nagar 400094, India}

\affiliation{\noindent $^{3}$Quantitative Life Sciences Section, The Abdus Salam International Centre for Theoretical Physics, 34151 Trieste, Italy}

\begin{abstract}
We introduce a spin-1 version of the random energy model with crystal field. Crystal field controls the density of $0$ spins in the system. We solve the  model in the micro-canonincal ensemble. The model has a spin-glass transition at a finite temperature for all strengths of the crystal field. By introducing the magnetic field we also obtain the de Almeida Thouless line for the model. The spin-glass transition persists in the presence of external field. We also find that the magnetisation shows non-monotonic behaviour for high positive crystal field strengths. The zero magnetic field specific heat and magnetic susceptibility also exhibit a cusp beyond a threshold value of the crystal field.
\end{abstract}

\date{\today}
\maketitle

\section{Introduction}
In a seminal work, an exactly solvable model of spin glasses, 
the random energy model (REM) was introduced by Derrida \cite{derrida}. Inspite of its simplicity, the model showed many features of spin-glasses like the freezing transition, a constant magnetic susceptibility in the spin-glass phase and the  persistence of spin-glass transition in magnetic field. Its generalisation, GREM also showed ultrametricity \cite{grem}. The model continues to be of interest as it provides an ideal platform to check for spin-glass theories \cite{derrida1,derrida2}. The energy in REM is a stochastic variable. The similar mathematical structure makes the model useful in the study of various other problems like error correcting codes \cite{dorlas,mezard}, deep learning \cite{marsili} and biological problems like protein folding \cite{protein}.

REM was originally defined as a model with Ising spins and uncorrelated energy levels, motivated by the $p \rightarrow \infty$ limit of the $p$-spin generalisation of the Sherrington-Kirkpatrick (SK) model. 

Ghatak and Sherrington\cite{ghatak} studied the spin-$1$ version of the SK model with a crystal field, also known as the random bond Blume-Capel model~\cite{ghatak,sherrington:22}. 
The $p \rightarrow \infty$ limit of the $p$-spin generalisation of this spin-$1$ disordered model was solved by Mottishaw \cite{mottishaw86}.
This model features a first order phase transition in virtue of the fact that, for any finite density of the $0$ spins, the random part of the energy due to $p$-spin interactions is sub-leading with respect to the crystal field. As a result, whenever the density of $0$ spins is non-zero, the energy is no longer a random variable and it equals the crystal field contribution to the energy of the $\pm 1$ spins. 

In this paper we define a different random energy model with spin-$1$ variables, where the energy contribution due to the clique containing the $\pm 1$ spins is assumed to be extensive. More precisely, for any configuration, the energy is assumed to be an i.i.d. Gaussian random variable with mean equal to the crystal field contribution and variance proportional to the number of $\pm 1$ spins. We call this model the crystal field random energy model (CREM). 

In general, disordered spin-$1$ models are known to exhibit inverse transitions. The re-entrance of order is observed in the studies of the spin-$1$ Ghatak-Sherrington model \cite{schupper,paoluzzi,leuzzi,costa} and also in experiments involving ferroelectric thin films, colloids, high-Tc superconductors, glasses and granular systems \cite{exp1,exp2,saratz2:10,brooke:99,rastogi:93,plazanet:04,angelini:08,avraham:01,sellitto:05}. It has also been observed in pure ferromagnetic Blume Capel model on hetergeneous graphs \cite{martino} and in the random field Blume Capel model \cite{santanu}.

Besides having a spin-glass transition, we shall show that the CREM does not exhibit a re-entrance, but rather it shows a crossover from a less ordered state to a more ordered state, depending on the crystal field, that is reminiscent to it.

The paper is organized as follows. We introduce the model in Sec. II. We solve the model in the micro-canonical ensemble in Sec. III.  In Sec. IV we solve the model in the presence of the magnetic field. We discuss our results in Sec. V

\section{Model} 

The REM for Ising spins was defined as a model with $2^N$ energy levels which are drawn from a probability distribution $P(E)$ \cite{derrida} that was taken to be a Gaussian with mean $0$ and variance $N/2$. Other choices for $P(E)$ have also been studied in the literature \cite{bouchaud}. 

%

In this spirit, here we consider a generalisation where the spin variables take three possible values i.e $s_i \in  \{\pm1,0\}$ $\forall i$, whereby spins $s_i=0$ do not contribute to the random part of the hamiltonian, which is assumed to be a Gaussian random variable with zero mean and variance $Q/2$, where $Q$ is the number of $\pm 1$ spins. As in Blume Capel models, we add to this random part a term $\Delta \sum_i s_i^2=\Delta Q$, where the crystal field $\Delta$ controls the density of the $0$ spins. Therefore, each state with $Q$ non-zero spins has an energy which is drawn i.i.d. from a distribution $P_Q(E) \propto e^{-{(E-\Delta Q)}^2/Q}$. The $\sum s_i^2$ term does not distinguish $+1$ and $-1$ spins, so the number of configurations for a given $Q$ are $2^Q {N \choose Q}$. The energy levels of this model are then drawn independently from the distribution
\begin{equation}
P(E) = \frac{1}{3^N} \sum_{Q=0}^N 2^Q {N \choose Q} \frac{e^{-(E-\Delta Q)^2/Q}}{C(Q)}
\label{El1}
\end{equation}
Requiring $P(E)$ to be normalised, gives $C(Q)=\frac{1}{\sqrt{\pi Q}}$.

We hence  define CREM as a spin-$1$ model with $3^N$  independent energy levels drawn from the energy distribution given by Eq. \ref{El1}.
We solve the model in the micro-canonincal ensemble.

\section{Entropy and the glass transition}

Let $n(E)dE$ be the random number that gives the number of energy levels in the interval $E$ and $E+dE$. Since the fluctutaions are small compared to the average value of $n(E)$ \cite{derrida}, $n(E) 
\sim \langle n(E) \rangle = 3^N P(E)$. Taking $\epsilon=E/N$ and $q=Q/N$, for large $N$, we replace the summation in the expression of $P(E)$ by an integral. We get
\begin{equation}
\langle n(E) \rangle  = \int_0^1 dq \exp{(N f(q,\epsilon,\Delta))}
\end{equation}
with  $f(q,\epsilon,\Delta)=q \ln 2 - q \ln q -(1-q) \ln(1-q) -\frac{(\epsilon- \Delta q)^2}{q}$.

\begin{figure}
\centering
\includegraphics[scale=0.6]{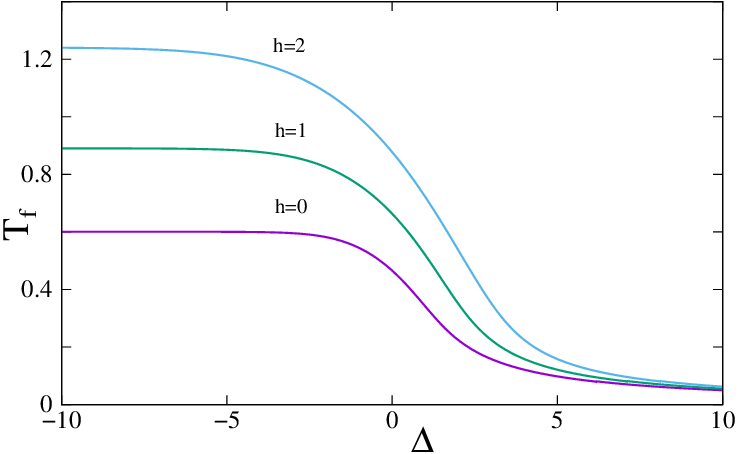}
\caption{Freezing transition temperature ($T_f(\Delta))$ as a function of the crystal field  ($\Delta$) for magnetic field strengths $h=0,1$ and $h=2$. The $h=0$ transition line corresponds of the three state REM in the absence of the external magnetic field.}
\label{TDpd}
\end{figure}

For large $N$ using the saddle point approximation we can write 
\begin{equation}
\langle n(E) \rangle  = \exp{(N f(q^*,\epsilon,\Delta))}
\end{equation}
where $q^*$ is the saddle point for a given $\epsilon$ and $\Delta$ and is given by the solution of the following equation
\begin{equation}
\ln 2 - \ln q^* + \ln (1-q^*) +\frac{(\epsilon-\Delta q^*)^2}{{q^*}^2} +\frac{2 \Delta (\epsilon-\Delta q^*)}{q^*} =0
\label{speq}
\end{equation}
The  log of the density of states is the entropy of the system. Hence we get the entropy ($S$) of the system as
\begin{equation}
\frac{S(\epsilon,\Delta)}{N}= -\frac{2 \epsilon^2}{q^*}+2 \epsilon \Delta -\log(1-q^*)
\label{entropyE}
\end{equation}
The values of $\epsilon$ and $q^*$ are related via the relation
\begin{equation}
|\epsilon| = q^* \sqrt{ \Delta^2 - \ln \frac{2 (1-q^*)}{q^*}}
\end{equation}
The entropy in Eq. \ref{entropyE} is positive only for a range of $\epsilon$. The entropy as a function of temperature ($T$) can also be calculated by using $\partial S/\partial \epsilon = \beta=1/T$. The entropy as a function of $T$ and $q^*$ is given by 
\begin{equation}
 \frac{S(T,\Delta)}{N} =\begin{cases}
    \frac{q^*}{T} \left(\Delta-\frac{1}{2T}\right)-\log(1-q^*) & \text{if $T > T_f(\Delta)$}.\\
   $0$ & \text{otherwise}.
  \end{cases}
  \label{entropy1}
\end{equation}
with 
\begin{equation}
q^* =\frac{2}{2+e^{(\frac{\Delta}{T}-\frac{1}{4 T^2})}}
\label{qfixed}
\end{equation}
and
\begin{equation}
\epsilon(T,\Delta) = q^* \left(\Delta-\frac{1}{2 T} \right)
\end{equation}
Equating $S(T,\Delta)=0$ gives the freezing/
glass transition temperature  $T_f(\Delta)$. The system freezes to the value $q^*_f(\Delta)$ at $T_f(\Delta)$ and the density of states is zero above this value of $q^*$.
As $\Delta$ increases $T_f(\Delta)$ decreases continuously (see Fig. \ref{TDpd}). For $\Delta \rightarrow -\infty$ model effectively has only $\pm 1$ spins. We find $T_f(\Delta) \rightarrow 0.6005$ as $\Delta \rightarrow -\infty$. This as expected is the freezing temperature of the Ising REM \cite{derrida}. For $\Delta=0$, the $T_f(\Delta) = 0.465454$ with $q^*_f = 0.8637$ and the  corresponding $|\epsilon| = 0.9279$. The variable $q^*$ freezes to a value $q^*_f(\Delta)$ for $T \leq T_f(\Delta)$. The entropy of the system goes to $0$ for $ T\le T_f(\Delta)$. 

From Eq. \ref{qfixed} it is clear that while $q^*$ is a montonically decreasing function of $T$ for $\Delta \leq 0$, it is a non-montonic function of $T$ for positive values of $\Delta$. It 
has a minimum at $T=1/2 \Delta$ for $\Delta>0$. The freezing temeperature  $T_f(\Delta)$ is strictly  less than  $1/2 \Delta$ for $\Delta>0$. To see this equate $S=0$ in Eq. \ref{entropy1}. This gives the value of $q^*$ at the freezing transition temperature to be
\begin{equation}
q^* = 1- \exp \left[\frac{q^*}{T_f(\Delta)} \left (\Delta-\frac{1}{2 T_f(\Delta)} \right) \right]
\end{equation}
Since $0 \le q^* \le 1$, it is clear that $T_f (\Delta) <1/2 \Delta$ for all the positive values of $\Delta$. The system hence freezes into a negative energy state with energy $\epsilon_f(T_f(\Delta),\Delta)= q^*(T_f(\Delta)) (\Delta-(1/2 T_f(\Delta))$ for $T<T_f(\Delta)$.

\begin{figure}
\centering
\includegraphics[scale=0.6]{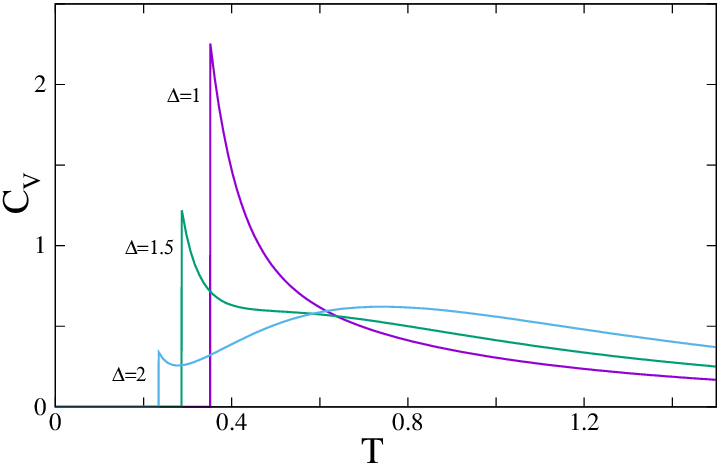}
\caption{Specific heat ($C_V$) in the absence of magnetic field ($h=0$) for $\Delta=1,1.5$ and $2$. It is $0$ below $T_f(\Delta)$. For $\Delta=1$ it decreases rapidly above $T_f(\Delta)$. For $\Delta=1.5$ it decays very slowly above the freezing transition temperature  and develops a cusp for $\Delta \gtrapprox 1.55$. We have plotted specific heat for $\Delta=2$ for which the cusp is clearly visible.}
\label{spheat}
\end{figure}

The free energy of the system comes out to be
\begin{equation}
 \frac{F(T,\Delta)}{N} =\begin{cases}
   T \log(1-q^*), & \text{if $T > T_f(\Delta)$}.\\
    q^*(T_f(\Delta)) \left(\Delta-\frac{1}{2 T_f(\Delta)}\right) & \text{otherwise}.
  \end{cases}
\end{equation}
The specific heat of the model shows an interesting behaviour. It is $0$ in the frozen phase and decays monotonically with increasing $T$ for  $T > T_f(\Delta)$ for $\Delta < 1.55$. But for $\Delta \gtrapprox 1.55$, it first decreases and then increases again peaking at a $T> T_f(\Delta)$ before decaying eventually (see Fig. \ref{spheat}).



\section{In the presence of the magnetic field}

\begin{figure}
\centering
\includegraphics[scale=0.6]{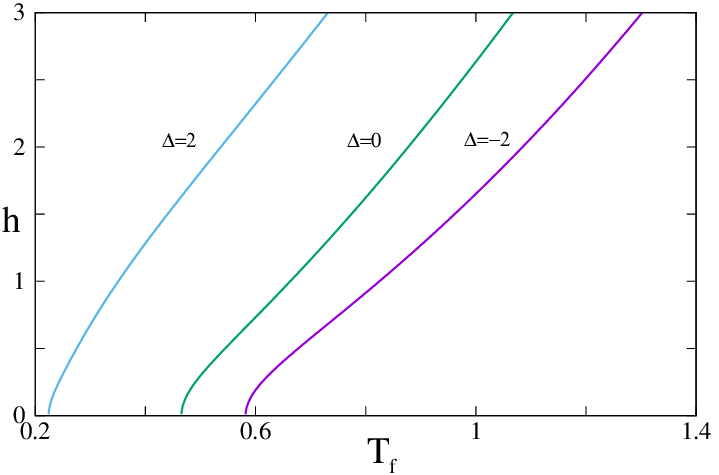}
\caption{$h-T_f$ transition lines for different values of the  crystal field ($\Delta$).}
\label{THpd}
\end{figure}
In order to study the magnetic properties we introduce a uniform magnetic field $h$ via a contribution $-h \sum_i s_i$ to the energy. The energy levels are again i.i.d. random variables with the distribution
\begin{equation}
P(E) = \frac{1}{3^N} \sum_{Q=0}^N {N \choose Q} \sum_{M=-Q}^{M=Q} {Q \choose (Q+M)/2} \frac{e^{-(E-\Delta Q+Mh)^2/Q}}{\sqrt{\pi Q}}
\end{equation}
here $M=Nm$ is the bulk magnetisation.

The average density  of states $\langle n(E)  \rangle = 3^N P(E)$. For large $N$ using the saddle point approximation we get the resultant entropy to be
\begin{equation}
\frac{S(\epsilon,\Delta)}{N} = q^* \ln \left( \frac{2 (1-q^*)}{\sqrt{{q^*}^2-m^2} } \right) -\frac{m}{2} \ln \left( \frac{q^*+m}{q^*-m} \right)- \ln  (1-q^*)
\end{equation}
where  $q^*$ and $m$ are functions of $\epsilon$ and $\Delta$ and are the solutions of the two simultaneous self-consistent equations
\begin{eqnarray}
y^2+2 \Delta y+\ln \left( \frac{2 (1-q^*)}{\sqrt{(q^*+m) (q^*-m)} } \right) &=& 0\\
\ln \left( \frac{q^*+m}{q^*-m} \right) -4 y h &=&  0
\end{eqnarray}
with  $y =  (\epsilon-\Delta  q^*+m h)/q^*$.

Using the above  two equations the entropy at a given energy can now be written as
\begin{equation}
\frac{S(\epsilon,\Delta)}{N} = \frac{2  \epsilon}{q^*} (\Delta q^*-\epsilon-m h) - \ln (1-q^*)
\end{equation}
It is more instructive to work with $T$ rather than $\epsilon$. We get the entropy as a function  of $T$ to be
\begin{equation}
\frac{S(T,\Delta)}{N} = \frac{q^*}{T} \left(\Delta-\frac{1}{2 T}-\frac{m h}{q^*}\right)-\ln(1-q^*)
\end{equation}
with \begin{eqnarray}
\label{qwh}
q^*& =&\frac{2}{2+ {\sech} (h/T) e^{(\frac{\Delta}{T}-\frac{1}{4 T^2})}}\\
\label{mwh}
m &=& q^* \tanh \frac{h}{T}
\end{eqnarray}
In this case also there is a critical value $T_f(h,\Delta)$ at which the entropy becomes zero. For $h=0$, the magnetisatio  $m=0$ for all non-zero $T$. For $h \neq 0$, $m$ is non-zero for any finite $T$. It freezes to a fixed value below $T_f (h,\Delta)$. But note that the transition is not in the spin-orientation. It is a freezing transition where the spins are frozen in a ferromagnetic state. Fig. \ref{TDpd}. shows the $T_f(h,\Delta)$ as a function of $\Delta$ for $h=1$ and $2$. Similar behaviour is seen for other values of $h$. We can also study the phase diagram in the $h-T$  plane. The line of phase transitions in the $h-T$ plane is known as the de Almeida Thouless line \cite{almeida}. The de Almeida Thouless line for different values of $\Delta$ is plotted in Fig. \ref{THpd}. For all values of $\Delta$ we find that the $T_f(h,\Delta)$ increases with $h$.

For $T \leq T_f(h,\Delta) \equiv T_f$, the system gets frozen in a zero entropy state with
\begin{eqnarray}
q_f^* &=&\frac{2}{2+ {\sech} (h/T_f) e^{(\frac{\Delta}{T_f}-\frac{1}{4 T_f^2})}}\\
m_f &=& q_f^* \tanh \frac{h}{T_f}
\end{eqnarray}
The energy is fixed at $\epsilon_f = q^*_f (\Delta-1/(2 T_f))-m_f h$ below $T_f$. For $T> T_f$ the energy is
\begin{equation}
\epsilon= q^* \left(\Delta-\frac{1}{2T}\right)-mh
\end{equation}
For $T> T_f$ the $q^*$ and $m$ values are given by equations Eq. \ref{qwh} and \ref{mwh} respectively. 

The free energy of the system is then given by
\begin{equation}
\frac{ F(T,h, \Delta)}{N} =\begin{cases}
   T \log(1-q^*), & \text{if $T > T_f$}.\\
    q^*_f (\Delta-\frac{1}{2 T_f}) - m_f h & \text{otherwise}.
  \end{cases}
\end{equation}

\begin{figure}
\centering
\includegraphics[scale=0.6]{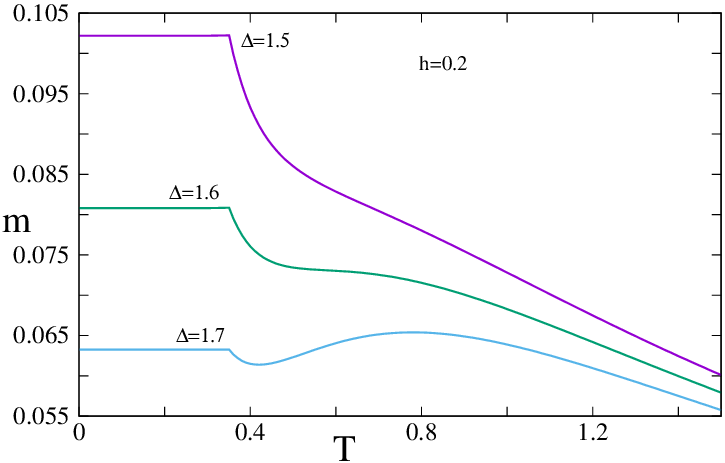}
\caption{$m-T$ plot for different values of $\Delta$ for $h=0.2$.}
\label{reentrance1}
\end{figure}

\begin{figure}[H]
     \centering
     \begin{subfigure}[b]{0.23\textwidth}
         \centering
         \includegraphics[width=\textwidth]{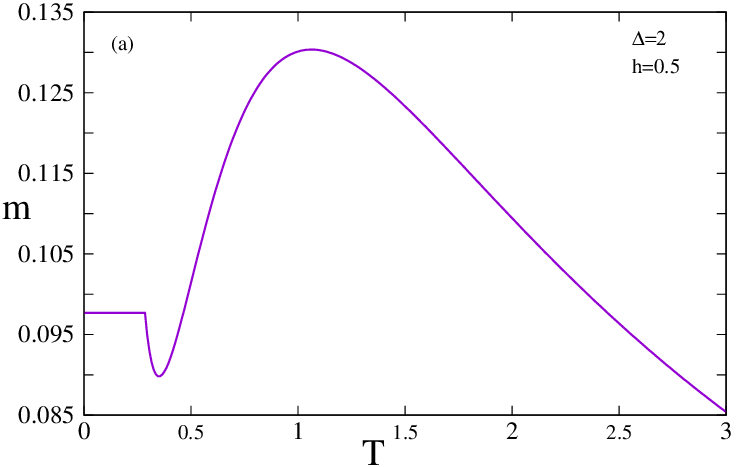}
         \label{fig:443}
     \end{subfigure}
     \begin{subfigure}[b]{0.23\textwidth}
         \centering
         \includegraphics[width=\textwidth]{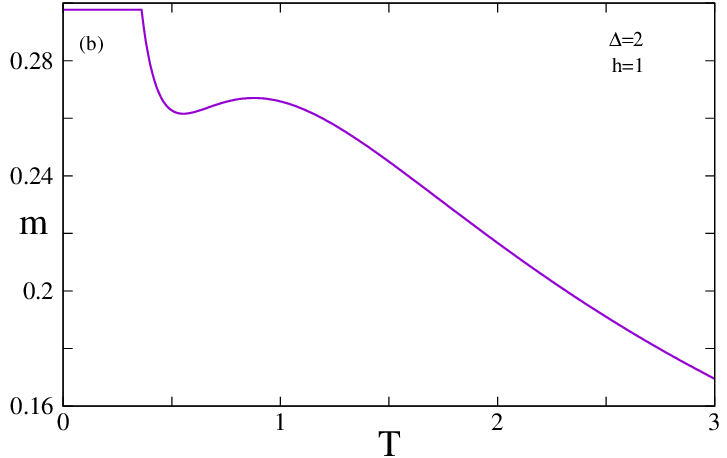}
         \label{fig:444}
     \end{subfigure}
     \hfill
     \begin{subfigure}[b]{0.23\textwidth}
         \centering
         \includegraphics[width=\textwidth]{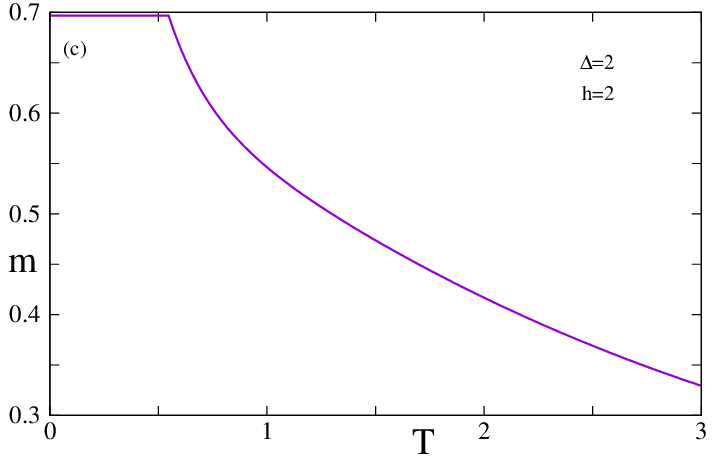}
         \label{fig:449}
     \end{subfigure}
     \hfill
     \begin{subfigure}[b]{0.23\textwidth}
         \centering
         \includegraphics[width=\textwidth]{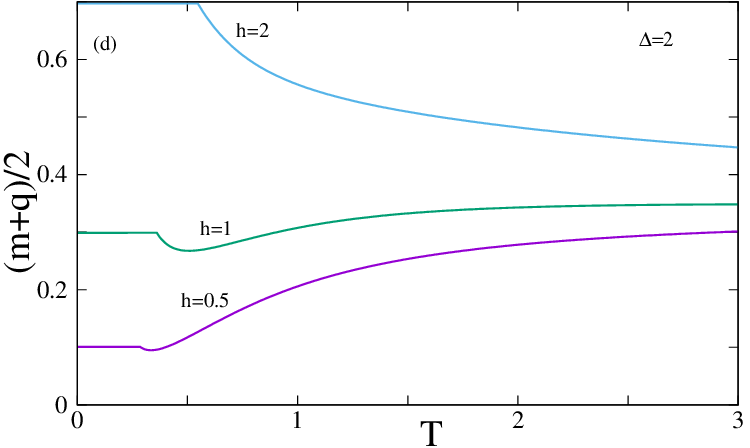}
         \label{fig:4410}
     \end{subfigure}
     \hfill
        \caption{(a),(b) and (c) show the $m-T$ plots for $h=0.5,1$ and $2$ respectively for $\Delta=2$. In (d) the average number of $+1$ spins as function of $T$  are plotted corresponding to $h$ and $\Delta$ values in (a),(b) and (c).}
        \label{reentrance2}
\end{figure}

For all values of $\Delta$ and $h$, first there is a freezing transition at $T_f(h,\Delta)$. Besides that for a fixed $h$ beyond a threshold value, i.e. for $\Delta > \Delta_{th}(h)$, as $T$ increases there is a non-monotonic increase of $m$ as shown in Fig. \ref{reentrance1}. For $\Delta > \Delta_{th}(h)$,  both $m$ and $q$ show non-montonic behaviour before approaching the values $0$ and $2/3$ respectively as $T \rightarrow  \infty$. The value $\Delta_{th}(h)$ increases with $h$. For example, as shown in the Fig. \ref{reentrance2}, for $\Delta=2$, while $m$ is non-monotonic for $h=0.5$ and $h=1$, it  decays monotonically for $h=2$. For $h=2$ the anamolous behaviour of $m$ is seen at a value of $\Delta >2$. 

For a fixed  value of $\Delta$, the non-monotonic behaviour of $m$ is present only for a finite range of $h$ (Fig. \ref{reentrance2}). To understand this we looked at the average number of $+1$ spins as a function of $T$ in the model. This is given by $(m+q)/2$. This value approaches $1/3$ as $T \rightarrow \infty$. We observe that when the  $(m+q)/2$ is less than $1/3$ at the freezing transition, there is a crossover to the higher ordered state as $T$ increases (see Fig. \ref{reentrance2}(d)). Below $T<T_f$ the system is frozen in a state with largely $0$ and $+1$ spins when $h \neq 0$. As $T$ increases it tries to reach an equal distribution over all three spin states. If $(m+q)/2 <1/3$ at $T_f$, the increase in the number of $+1$ spins is more rapid than the increase in the number of $-1$ spins, giving rise to peak in $m$.

\begin{figure}
\centering
\includegraphics[scale=0.6]{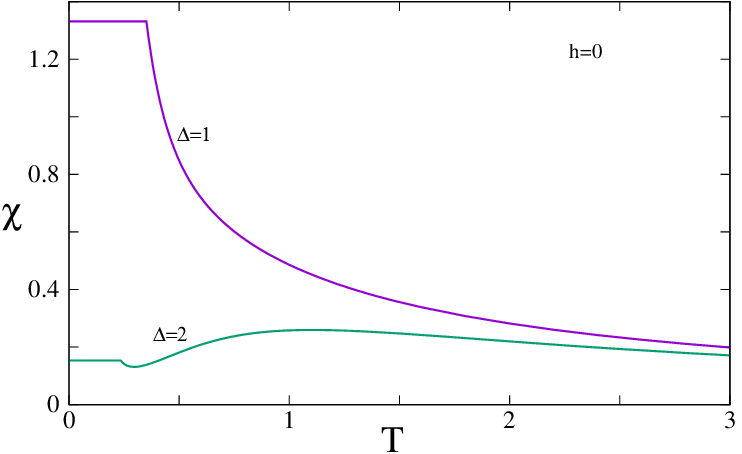}
\caption{Zero magentic field magnetic susceptibility ($\chi$) for $\Delta=1$ and $2$.}
\label{chi}
\end{figure}

We can also calculate the zero field magnetic susceptibility($\chi$) . It is
\begin{equation}
\chi =\begin{cases}
  \frac{q^*_f}{T_f}, & \text{if $T < T_f(0,\Delta)$}.\\
    \frac{q^*}{T} & \text{otherwise}.
  \end{cases}
\end{equation}
Below the glass transition, even though there is no magnetisation, the zero field magnetic susceptibility is a  constant. It decreases above the glass transition for small values of $\Delta$, but for large positive $\Delta$ it shows a cusp just like the zero field specific heat. We have plotted it for $\Delta=1$ and $2$ in Fig. \ref{chi}.

\section{Discussion}

Introduction of quenched disorder is known to give rise to new and unexpected phenomenon compared to  the pure models, like the existence of the spin glass phase and re-entrance of order at higher $T$. The exactly solvable CREM introduced in this paper, generalises the original REM to systems with two order parameters. While the original REM shows a spin glass phase, we find that the introduction of a crystal field induces a crossover to a higher magnetised state at higher $T$. Unlike the Ghatak-Sherrington model, where the re-entrance is associated with a phase transition, the non monotonic behaviour of the order parameter in CREM is a simple crossover. It would be interesting to find the conditions that can induce a true inverse transition in the model.  No re-entrance was reported in the $p \rightarrow \infty$ limit of the Ghatak-Sherrington model \cite{mottishaw86}. A recent work shows that the introduction of a quenched random field from an asymmetric distribution introduces inverse transitions in the random-field ferromagnetic spin-1 models \cite{santanu}. It would be interesting to study the effects of quenched random magnetic field \cite{filho06,arguin14} and crystal fields in CREM. A generalisation of the quantum random energy model (QREM) \cite{qrem} to a model with triplet spins and crystal field can also be studied. It would also be useful to solve the CREM in the canonical ensemble and also using the theory of large deviations.


{\it Acknowledgement : }
S would like to acknowledge support from the ICTP through the Associates Programme and from the Simons Foundation through grant number 284558FY19.

\end{document}